\documentclass[twocolumn,superscriptaddress]{revtex4-2}
\usepackage[english]{babel}
\usepackage[colorinlistoftodos, color=green!40, prependcaption]{todonotes}
\usepackage{amsthm}
\DeclareUnicodeCharacter{2212}{-}
\usepackage{mathtools}
\usepackage{physics}
\usepackage{xcolor}
\usepackage{graphicx}
\usepackage{comment}
\usepackage{ulem}
\usepackage[left=23mm,right=13mm,top=35mm,columnsep=15pt]{geometry} 
\usepackage{lipsum}
\usepackage{csquotes}
\usepackage[pdftex, pdftitle={Article}, pdfauthor={Author}]{hyperref}
\usepackage[dvipsnames]{xcolor}
\usepackage{bm}
\usepackage{siunitx}
\usepackage{xr}
\usepackage{comment}
\usepackage{soul}

\addto\captionsenglish{}

\begin{document}
\graphicspath{{./Figures/}{./}}
%\preprint{APS/123-QED}

\title{Wave-based reading of mechanical memory in multistable mass-in-mass metamaterials}

\author{Audrey A. Watkins}
\affiliation{John A. Paulson School of Engineering and Applied Sciences, Harvard University, Cambridge, MA 02138, USA}

\author{Giovanni Bordiga}
\affiliation{John A. Paulson School of Engineering and Applied Sciences, Harvard University, Cambridge, MA 02138, USA}

\author{Vincent Tournat}
\affiliation{Laboratoire d'Acoustique de l'Universit\'e du Mans, UMR 6613, Institut d'Acoustique -- Graduate School, CNRS, Le Mans Universit\'e, Le Mans, France}
\affiliation{John A. Paulson School of Engineering and Applied Sciences, Harvard University, Cambridge, MA 02138, USA}

\author{Katia Bertoldi}
\affiliation{John A. Paulson School of Engineering and Applied Sciences, Harvard University, Cambridge, MA 02138, USA}

\date{\today}

\begin{abstract}
Mechanical metamaterials with integrated bistable elements have emerged as promising platforms for mechanical information storage, where transitions between stable states encode information as mechanical bits. While it has been shown that information can be   written into such metamaterials by applying global inputs, existing readout strategies rely predominantly on visual inspection. Here, we experimentally demonstrate a mass-in-mass bistable metamaterial with state-dependent stiffness that enables both writing and reading of mechanical information using only boundary-applied dynamic excitations. We further show that such metamaterial architecture functions as both a mechanical sensor of input amplitude and a reconfigurable wave-control device. Together, these results highlight the versatility of bistable metamaterials as multifunctional platforms that integrate mechanical memory, sensing, and adaptive wave manipulation.

\end{abstract}

\maketitle

\section{Introduction}

Mechanical metamaterials are artificially structured systems whose geometry, rather than material composition alone, governs their atypical and tunable mechanical properties \cite{bertoldi2017flexible, jiao2023mechanical, barchiesi2019mechanical, deng2021nonlinear}. Through careful geometric design, numerous advanced capabilities have been realized including programmable shape morphing~\cite{dudek2025shapemorphingmetamaterials,jin2020kirigami,Ni2022}, energy trapping~\cite{shan2015multistable}, and wave guiding and filtering~\cite{deymier2013acoustic,DORN2023102091,CelliGonella}. In particular, mechanical metamaterials composed of bistable elements have emerged as promising platforms for storing and processing mechanical information \cite{chen2021reprogrammable, watkins2025arbitrary, yasuda2017origami, pechac2023mechanical, gutierrez2026dynamic, WU2026, kwakernaak2023counting}. Arrays of multistable elements, such as bistable beams, shells, and trusses, can support interesting wave phenomena such as transition waves \cite{jin_transition_2026, vasios_universally_2021, raney_stable_2016, jin2020guided, Hwang2018}, and can also be ``programmed" into a variety of stable configurations by applying specific inputs \cite{pechac2023mechanical, watkins2025arbitrary, gutierrez2026dynamic, liu2024controlled, Wu2026_advmat, jiao2024phase}. In this context, the act of switching bistable elements between their stable states is viewed as \textit{writing} information into the structure, analogous to programming digital bits. While most methods of writing are quasi-static and require localized inputs to reconfigure each individual bistable element \cite{chen2021reprogrammable, zhang2025inmemoryphononiclearningcognitive, udani_programmable_2021, tahidul_haque_reprogrammable_2024, Wu2026_advmat}, recent work has shown this writing can be achieved by using one global dynamic input to trigger the transition of one or multiple bistable elements at once \cite{gutierrez2026dynamic,watkins2025arbitrary}.

Thus far, the \textit{read-out} of information encoded in multistable mechanical metamaterials has typically relied on visually identifying the state of each bistable element \cite{yasuda2017origami,watkins2025arbitrary,gutierrez2026dynamic}. However, there is a growing need for read-out strategies that do not require visual inspection and can instead infer the state of the metamaterial by monitoring its response to a prescribed input. Initial steps in this direction have been demonstrated using both the quasi-static response of the system under loading \cite{chen2021reprogrammable} and its dynamic response as a read-out mechanism \cite{pechac2023mechanical,zhang2025inmemoryphononiclearningcognitive}.

Building on these advances, we show that a mass-in-mass mechanical metamaterial, previously demonstrated to enable the arbitrary dynamic writing of mechanical information through large-amplitude boundary pulses \cite{watkins2025arbitrary}, can also support dynamic readout by introducing state-dependent stiffness into its bistable elements. We show that this read-out strategy naturally extends to larger arrays and can be harnessed for functionalities such as wave control and sensing mechanical pulses. Importantly, our approach operates entirely in the dynamic regime: the global quasi-static properties of the metamaterial remain unchanged regardless of the information encoded within it. 

The paper is organized as follows. In Section~\ref{sec:Mass-in-mass metamaterial with state-dependent stiffness}, we introduce and characterize the mass-in-mass metamaterial with state-dependent stiffness. In Section~\ref{sec:Writing and reading arbitrary mechanical memory in an $N=3$ structure}, we demonstrate how information can be dynamically written into the structure using large-amplitude boundary pulses and subsequently read out using low-amplitude pulses. Finally, in Sections~\ref{sec:Sensing excitation amplitudes in an $N=8$ structure}  and ~\ref{sec:Wave control in an $N=8$ structure} we extend these capabilities to a metamaterial comprising a larger number of unit cells and demonstrate enhanced sensing and wave-control functionalities.

% Motivated by these advances, in this work we experimentally demonstrate that mass-in-mass mechanical metamaterial, previously shown to enable the arbitrary dynamic writing of mechanical information through the application of large-amplitude pulses at its boundary \cite{watkins2025arbitrary}, \VT{can be enhanced/adapted/enriched to also} support dynamic read-out through the application of low-amplitude probing signals \VT{to be discussed}. This capability is achieved by introducing state-dependent   stiffnesses for the bistable elements. We show that this read-out strategy naturally extends to larger arrays and can be harnessed for functionalities such as wave control and \VT{sensing mechanical pulses}. Importantly, our approach operates entirely in the dynamic regime: the global quasi-static properties of the metamaterial remain unchanged regardless of the information encoded within it.
% \VT{What about a sentence or two introducing what we do in the following, the article structure?} 

\section{Mass-in-mass metamaterial with state-dependent stiffness}
\label{sec:Mass-in-mass metamaterial with state-dependent stiffness}

\begin{figure}{!hb}
\centering
\includegraphics[width=1\linewidth]{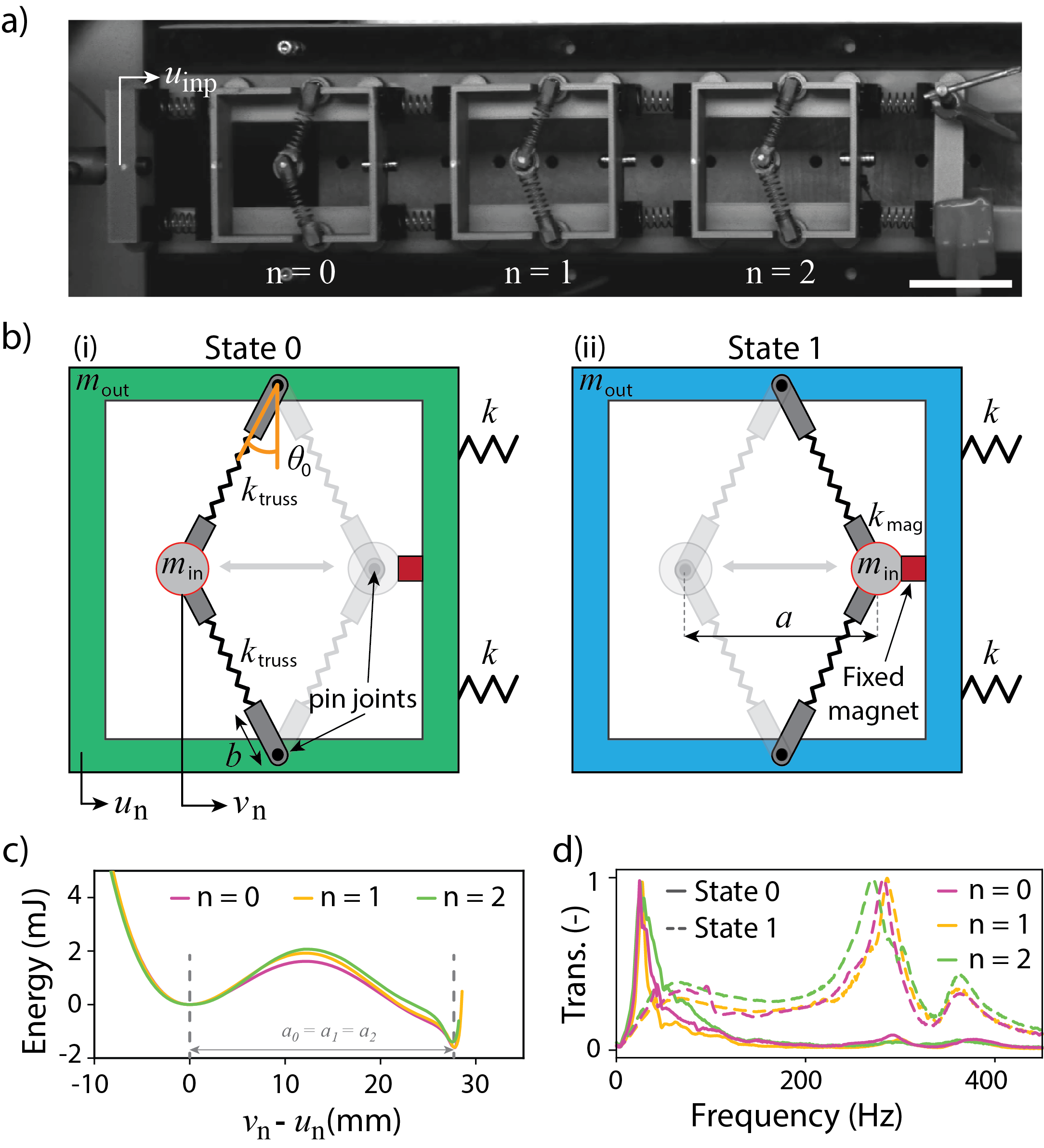}
\caption{\textbf{Mass-in-mass metamaterial with state-dependent stiffness.} (a) Photograph of a mass-in-mass metamaterial with $N=3$ unit cells. Scale bar: 5 mm. (b) Schematic of a mass-in-mass unit cell in (i) \textit{state 0} and (ii) \textit{state 1}. (c) Experimentally measured asymmetric energy landscapes of the bistable elements for the three unit cells. (d) Measured frequency response of each unit cell in its two stable configurations.}
\label{fig:figure 1}
\end{figure}

Our mass-in-mass metamaterial consists of a one-dimensional array of $N$ elastically coupled unit cells, each containing a bistable element (Fig.~\ref{fig:figure 1}a shows a system with $N=3$). Each unit cell is composed of an outer mass, $m_{\text{out}}=33.6$ g, and an inner mass, $m_{\text{in}}=3.8$ g, coupled through a bistable von Mises truss (Fig.~\ref{fig:figure 1}b). The outer masses of adjacent unit cells are further coupled through two linear elastic springs in parallel, each with stiffness $k=1020$ N/m. The bistable energy landscape of the von Mises truss depends on the stiffness, $k_{\text{truss}}$, and the initial angle, $\theta_0$, of the two springs forming the truss. The springs are connected to the outer mass through a rigid link of length $b$, with near-frictionless pin joints at each connection. By varying $b$, we tune the initial inclination $\theta_0$ of the truss members and, consequently, the energy barrier height $\mathcal{E}$ and general shape of the bistable potential (Fig.~\ref{fig:figure 1}c). As recently shown \cite{watkins2025arbitrary}, such bistable elements that possess two stable configurations, namely \textit{state 0} and \textit{state 1}, with symmetric energy landscapes can enable arbitrary mechanical information to be encoded through nonlinear waves applied at the boundary of the system.

Here, instead, to facilitate information readout through small-amplitude pulses applied at the boundary, we intentionally introduce stiffness asymmetry between the two stable states, resulting in distinct elastic responses of the bistable element in each configuration. This asymmetry is achieved by attaching a permanent magnet to one side of the outer mass, such that it interacts with the inner mass only in one of the two stable states. For the three unit cells considered in Fig.~\ref{fig:figure 1}, the  distance between the inner mass and the magnet is experimentally measured to be $(a_0,~a_1,~a_2)=(27.7~\text{mm},~27.8~\text{mm},~27.6~\text{mm})$ in \textit{state 0} and $0$ mm in \textit{state 1}. Consequently, in \textit{state 1}, $m_{\mathrm{in}}$ comes into contact with the magnet rigidly attached to $m_{\mathrm{out}}$. The resulting attractive interaction modifies the bistable energy landscape, introducing asymmetry between the two local energy minima (Fig.~\ref{fig:figure 1}c). Moreover, the associated change in local stiffness produces a state-dependent shift in the resonance frequency of each inner mass. To quantify this shift, each unit cell is excited using a low-amplitude chirp signal spanning $f\in[3,\,450]$ Hz, and the average resonance frequencies are identified as $\omega_0 = 26$ Hz and $\omega_1 = 280$ Hz from the fast Fourier transform of the measured dynamic response (Fig.~\ref{fig:figure 1}d). In the following sections, we demonstrate that introducing state-dependent stiffness enables both the dynamic writing and dynamic readout of mechanical information in the metamaterial.

\section{Writing and reading arbitrary mechanical memory in an $N=3$ structure}
\label{sec:Writing and reading arbitrary mechanical memory in an $N=3$ structure}

\begin{figure}
\centering
\includegraphics[width=1\linewidth]{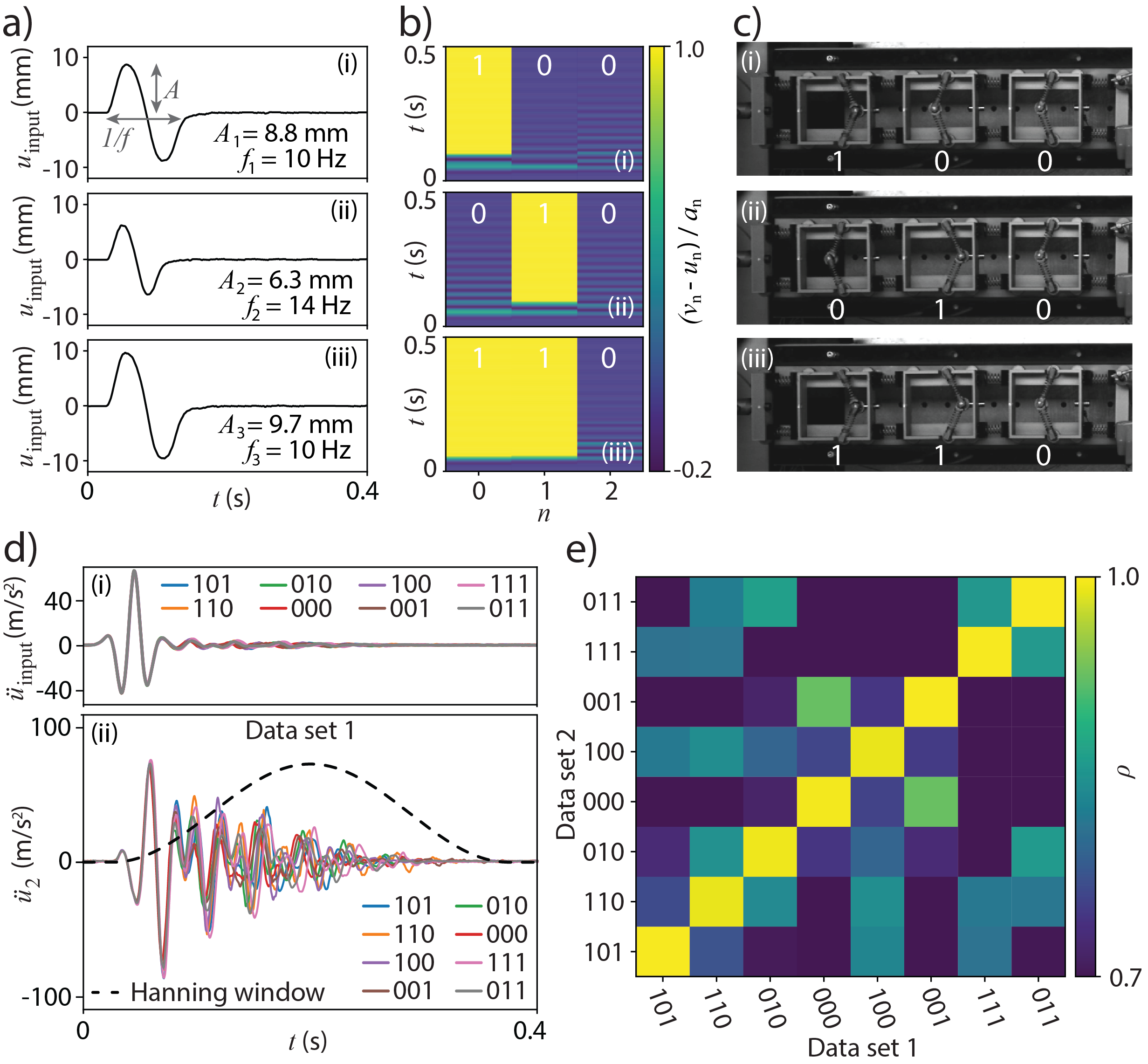}
\caption{\textbf{Dynamic writing and readout.} (a) Experimentally recorded large-amplitude input signals applied to the left boundary of the metamaterial. (b) Corresponding spatio-temporal response of the metamaterial. (c) Images of three  final configurations at $t=0.5~\mathrm{s}$. (d) (i) Nominally identical low-amplitude probing signals applied to the left boundary of the metamaterial and (ii) the corresponding measured acceleration of the $n=2$ unit, $\ddot{u}_2(t)$, for each of the $2^N$ stable configurations supported by the metamaterial. The dotted line indicates the Hanning window used for the cross-correlation analysis. (e) Maximum normalized cross-correlation coefficient, $\rho$, between a reference data set (Data set 1) and a second data set (Data set 2) for all $2^N$ configurations.}
\label{fig:figure 2}
\end{figure}

As for the metamaterial with bistable elements possessing a symmetric energy landscape \cite{watkins2025arbitrary}, the state of the $N=3$ metamaterial with bistable elements exhibiting an asymmetric energy landscape can be dynamically written and controlled by applying a large-amplitude input at the boundary of the structure (Fig.~\ref{fig:figure 2}a). The large amplitude inputs cause oscillatory motion of both $m_{\text{in}}$ and $m_{\text{out}}$ in all three unit cells due to their couplings. If the oscillations are sufficiently large, one or more bistable elements overcome their energy barriers and transition to \textit{state 1}, where they come into contact with rigid magnets that latch them into this state (Fig.~\ref{fig:figure 2}b). The specific final configuration reached under a given input results from highly complex interactions among all degrees of freedom, with the sensitivity to variations in the input parameters being strongly influenced by the unit-cell geometry \cite{watkins2025arbitrary}. Consequently, by appropriately selecting the input amplitude and frequency, the metamaterial can be reconfigured into arbitrary target states \cite{watkins2025arbitrary}.

To investigate this behavior experimentally, one end of the sample is fixed while a low-frequency shaker applies a bipolar pulse to the opposite end. The pulse is generated from a single period of a sinusoidal electrical signal and is characterized by an amplitude $A$ and frequency $f$. The response of the metamaterial is recorded using a high-speed overhead camera, and the displacements of all inner and outer masses are extracted through point tracking. Figs.~\ref{fig:figure 2}a--c show three representative experiments in which the metamaterial is initially prepared in state $000$, corresponding to all three unit cells residing in \textit{state 0}. Three distinct input pulses are applied, each characterized by a sinusoidal signal with measured amplitude and prescribed frequency $(A_1,f_1)=(8.8~\mathrm{mm},10~\mathrm{Hz})$, $(A_2,f_2)=(6.3~\mathrm{mm},14~\mathrm{Hz})$, and $(A_3,f_3)=(9.7~\mathrm{mm},10~\mathrm{Hz})$. For each experiment, we track the evolution of the relative displacement of the $n$th unit cell, $v_n(t)-u_n(t)$, where $v_n$ and $u_n$ denote the horizontal displacements of the inner and outer masses, respectively. As shown in Fig.~\ref{fig:figure 2}b, the applied excitations drive one or more bistable elements across their energy barrier, leading to transitions into \textit{state 1}. Specifically, the three input pulses reconfigure the metamaterial from the initial state $000$ to the final states $100$, $010$, and $110$, respectively (Fig.~\ref{fig:figure 2}c).

Due to the state-dependent local stiffnesses, the low-amplitude dynamic responses of the sample are altered as the sample is reconfigured into any of the $2^N$ possible configurations. To experimentally observe this, we use a low-frequency shaker (The Modal Shop$^{\copyright}$ model 2025E) to apply nominally identical low-amplitude Ricker wavelet input signals characterized by a central frequency $f_0=30$ Hz, peak amplitude $\Bar{A}=1.8$ mm, and -3 dB bandwidth of 23.3 Hz (Fig. \ref{fig:figure 2}d(i)). We then monitor the temporal evolution of the acceleration in the longitudinal direction of the outer mass of unit cell $n=N-1=2$, $\ddot{u}_2(t)$,  using a single-axis accelerometer (PCB Piezotronics model Y352C23) attached to the mass. Fig.~\ref{fig:figure 2}d(ii) presents the  evolution of $\ddot{u}_2(t)$, averaged over two trials for each configuration. Because each configuration is associated with a distinct distribution of local stiffnesses, the low-amplitude responses differ in both amplitude and phase, suggesting a pathway for reliable distinction among the $2^N$ states.

The data presented in Fig.~\ref{fig:figure 2}d(ii) - data set 1 - can act as a reference experimental library of data to be compared against secondary experimental data sets. To evaluate the suitability of data set 1 acting as the reference, we perform and record the data for a second set of experiments - data set 2 - and perform a cross correlation analysis to compute the maximum normalized cross correlation coefficient $\rho$ between the temporal signals corresponding to each of the $2^N$ configurations of the metamaterial. Fig.~\ref{fig:figure 2}e highlights the strong agreement between temporal signals in the reference library - data set 1 - and a secondary data set 2, as indicated by the diagonal values that are close to unity, allowing us to effectively dynamically read which configuration the sample rests in solely through a cross-correlation analysis.

\begin{figure}
\centering
\includegraphics[width=1\linewidth]{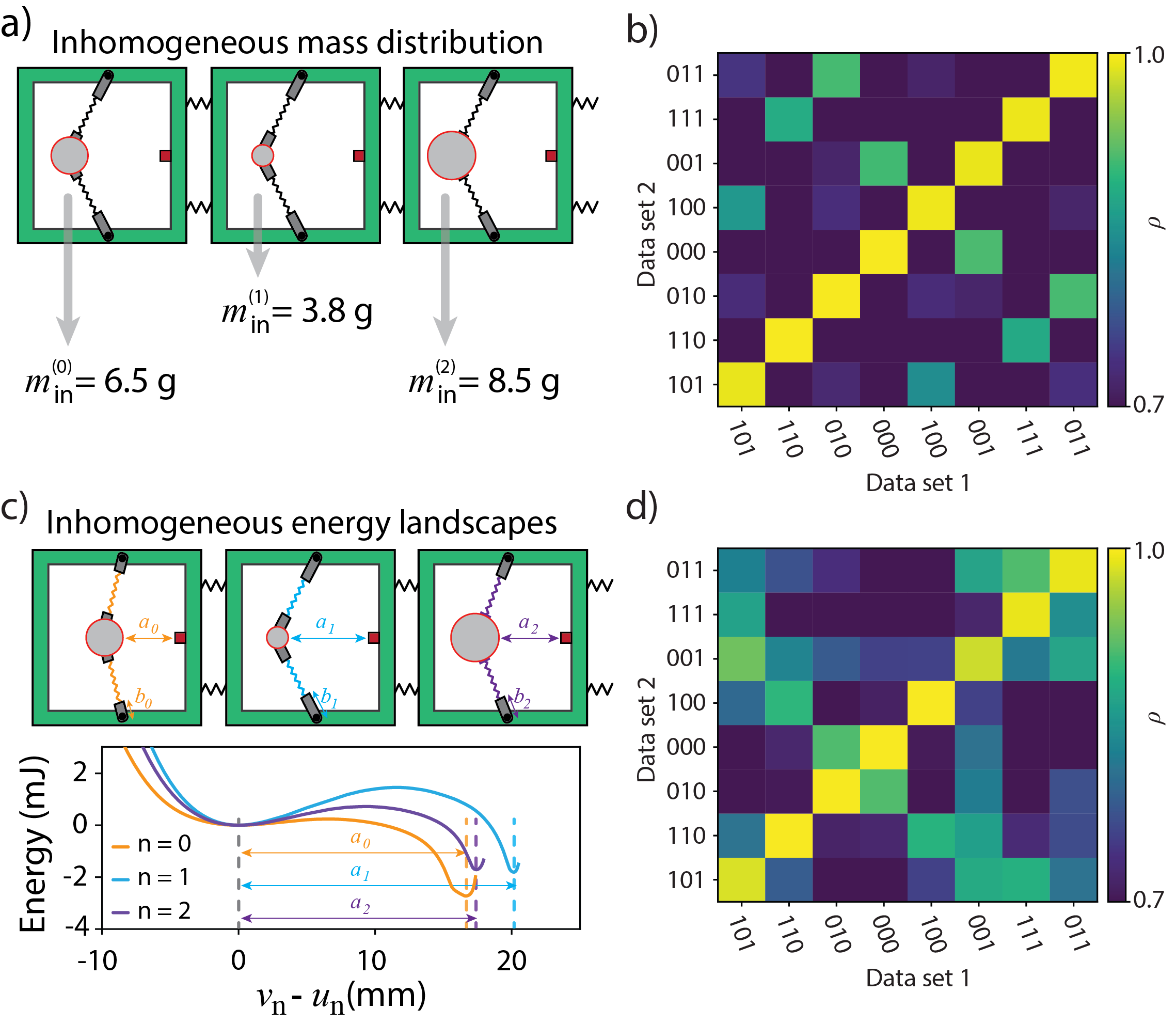}
\caption{\textbf{Metamaterials with inhomogeneous mass and bistable energy landscapes.} (a) Schematic of a metamaterial with three unit cells having different inner masses, $m_{\mathrm{in}}^{(n)}$. (b) Corresponding cross-correlation analysis of the low-amplitude responses between two independent experimental data sets. (c) Schematic of a metamaterial with three bistable elements exhibiting different energy landscapes and different inner masses. (d) Corresponding cross-correlation analysis of the low-amplitude responses between two independent experimental data sets.}
\label{fig:figure 3}
\end{figure}

Up to this point, we have restricted our attention to systems composed of nominally identical unit cells. However, such uniformity is not required for our approach: the individual unit cells can be varied independently while preserving the same design principles and functionality. To demonstrate this capability, we modify the inner masses of the three unit cells to $m_{\mathrm{in}}^{(0)}=6.5$ g, $m_{\mathrm{in}}^{(1)}=3.8$ g, and $m_{\mathrm{in}}^{(2)}=8.5$ g, as shown in Fig.~\ref{fig:figure 3}a. We then reconfigure the structure into each of the $2^N$ possible configurations and apply low-amplitude input signals that are nominally identical to those used in Fig.~\ref{fig:figure 2}d(i). We find that the system configuration can again be uniquely identified by computing the cross-correlation between sets of measured low-amplitude responses (Fig.~\ref{fig:figure 3}b, see also Supplementary Section II). Similar results are obtained when, in addition to varying the inner masses, the bistable energy landscape of the $n$th unit cell is modified by changing the length of the rigid pin-jointed struts, $b_n$. This variation alters both the initial truss angle, $\theta_0^{(n)}$, the distance between the stable states, $a_n$, and the energy barrier height, $\mathcal{E}_n$ (Fig.~\ref{fig:figure 3}c). In Fig.~\ref{fig:figure 3}d we present results for a metamaterial with ($a_0,b_0,\mathcal{E}_0$)=($16.5~\mathrm{mm}$, $8.3~\mathrm{mm}$, $0.19~\mathrm{mJ}$), ($a_1,b_1,\mathcal{E}_1$)=($20.1~\mathrm{mm}$, $9.3~\mathrm{mm}$, $1.37~\mathrm{mJ}$), and ($a_2,b_2,\mathcal{E}_2$)=($17.4~\mathrm{mm}$, $8.7~\mathrm{mm}$, $0.67~\mathrm{mJ}$). We again probe the structure using low-amplitude inputs that are nominally identical to those employed in Fig.~\ref{fig:figure 2}e and record the corresponding responses (Fig.~\ref{fig:figure 3}d).
Interestingly, the resulting cross-correlation matrix exhibits slightly larger off-diagonal values of $\rho$ than those observed for the previous examples, indicating greater similarity among some of the measured $\ddot{u}_{2}$ in response to the low-amplitude Ricker wavelet. Nevertheless, all eight configurations remain distinguishable, demonstrating that reliable state identification is preserved even when both the masses and the bistable energy landscapes are varied across the structure. %\AW{In the next sections, we illustrate how this read-out process can be extended to larger samples, and we highlight the emergent sensing and wave control functionalities that result.}

\section{Sensing excitation amplitudes in an $N=8$ structure}
\label{sec:Sensing excitation amplitudes in an $N=8$ structure}

\begin{figure*}
\centering
\includegraphics[width=1\linewidth]{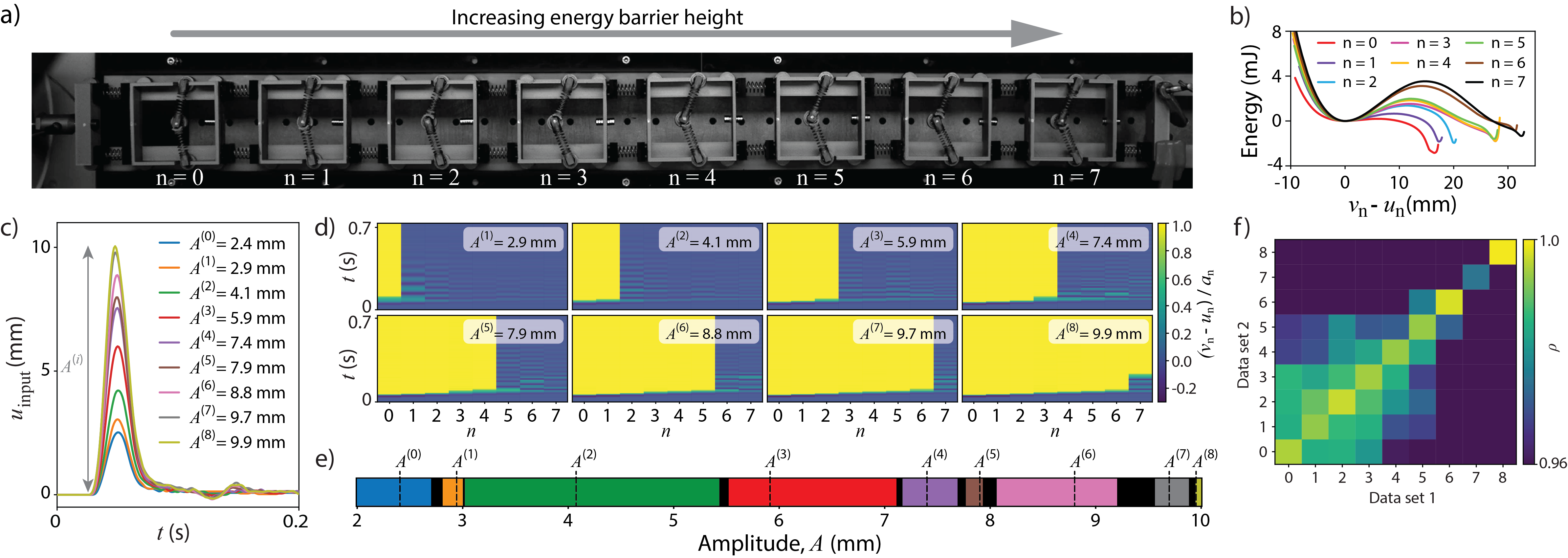}
\caption{\textbf{Sensing excitation amplitudes in an $N=8$ sample.} a) Photographed $N=8$ sample and the corresponding b) experimentally measured bistable energy landscapes. Experimentally recorded dynamic c) input signals and d) normalized sample responses. e) Amplitude sensing schematic with regions of sensitivity indicated (black). f) Cross-correlation analysis of the low amplitude responses between two sets of experimental data with axis labels corresponding to the number of unit cells in \textit{state 1}.}
\label{fig:figure 4}
\end{figure*}

So far, we have used large-amplitude pulses applied at the boundary to drive the metamaterial into an arbitrary state, and small-amplitude pulses to read out such state. We now demonstrate that these capabilities can be leveraged to sense features of an applied large-amplitude excitation. Specifically, our goal is to design a structure whose final state uniquely encodes the amplitude of the incoming excitation. Towards this end, we consider a chain comprising $N=8$ unit cells, whose boundary is subjected to a large-amplitude single-haversine displacement pulse of frequency $\Bar{f}$ and varying amplitude $A\in[2,10]$ mm. By examining the resulting final configuration of the structure, we seek to directly infer the amplitude of the applied excitation.

To achieve this, we design the structure such that the unit cells exhibit a spatial gradient in energy barrier heights, with unit cells ($n=0$) and ($n=7$) requiring the least and greatest amount of energy, respectively, to transition from \textit{state 0} to \textit{state 1} (Fig. \ref{fig:figure 4}a,b). This design is expected to favor the switching of cells with lower energy barriers, thereby enabling a sequential activation process in which progressively larger excitation amplitudes induce transitions in cells with increasingly higher energy barriers. In contrast, if all unit cells shared identical energy barriers, prior work suggests that transitions would occur in a non-sequential manner (e.g., (00000000 $\rightarrow$ 00100000)), resulting in a less direct relationship between the final state and the excitation amplitude \cite{watkins2025arbitrary}.

We tune the energy landscapes by varying $b_n$ such that ($b_0$, $b_1$, $b_2$, $b_3$, $b_4$, $b_5$, $b_6$, $b_7$) = ($8.3~\mathrm{mm}$, $8.7~\mathrm{mm}$, $9.3~\mathrm{mm}$, $10.0~\mathrm{mm}$, $10.9~\mathrm{mm}$, $11.0~\mathrm{mm}$, $11.9~\mathrm{mm}$, $12.1~\mathrm{mm}$). As shown in Fig. \ref{fig:figure 4}b, by modifying $b_n$, both the second equilibrium position $a_n$ and energy barrier height $\mathcal{E}_n$ are modified (see Table SI). To ensure contact between the rigid magnet and the inner mass in \textit{state 1}, we adjust the length of the magnet interacting with the inner mass to accommodate the different values of  $a_n$.

In the experiments, we fix one end of the sample and apply a single-haversine displacement pulse of frequency $\bar{f}=12$ Hz and varying amplitude $A$ at the opposite end. Fig.~\ref{fig:figure 4}d presents the experimentally observed responses of the sample for the nine inputs shown in Fig.~\ref{fig:figure 4}c, as recorded by an overhead high-speed camera (see also Video S1). At sufficiently small amplitudes, such as $A=A^{(0)}=2.4$ mm ($A^{(m)}$ denoting an input amplitude for which $m$ unit cells transition from \textit{state 0} to \textit{state 1}$)$, none of the eight unit cells transition from \textit{state 0} to \textit{state 1}. As the amplitude increases, state transitions begin to occur. For example, at $A=A^{(1)}=2.9$ mm, unit cell $n=0$ transitions to \textit{state 1}. Increasing the amplitude further to $A=A^{(2)}=4.1$ mm causes both unit cells $n=0$ and $n=1$ to transition. With further increases in $A$, a transition wave front propagates progressively deeper into the sample, with the penetration depth directly correlated with the input amplitude.

% we perform roughly 300 experiments, finely sweeping the input amplitude between 2 mm to 10 mm by incrementally increasing the drive voltage of the function generator by $3$ mV, resulting in $\approx3~\mu$m incremental increase in input amplitudes, and identify the resulting final configuration of the structure}. 

To systematically characterize this behavior, we finely sweep the input amplitude by incrementally increasing the drive voltage of the function generator in steps of $3$ mV, corresponding to an approximately $3~\mu$m increase in the measured amplitude of the applied pulse (see Supplementary Information Section II for the voltage-to-displacement calibration). We perform approximately 300 experiments while varying the measured input amplitude between 2 mm and 10 mm and note the resulting final configuration of the metamaterial. As shown in Fig.~\ref{fig:figure 4}e, this procedure enables us to identify amplitude intervals over which the system consistently converges to each of the nine configurations, as well as narrow transition regions in which small changes in the input amplitude can lead to different final states.

Importantly, as demonstrated  for the metamaterial with $N=3$ unit cells, the state of the longer $N=8$ chain can also be read out by applying a small-amplitude pulse. Fig.~\ref{fig:figure 4}f shows that the state of the metamaterial can be successfully identified by performing a cross-correlation between a stored response corresponding to a known configuration (data set 1) and a subsequently measured response from a sample in an unknown configuration (data set 2). Furthermore, because the number of switched unit cells is strongly correlated with the amplitude of the applied dynamic input, the configuration readout can also be interpreted as a means of identifying the input amplitude. In this sense, the metamaterial functions as a coarse-grained, discrete amplitude sensor: the final configuration encodes the amplitude of the applied haversine pulse, while the low-amplitude readout response provides a means to retrieve this information.

\section{Wave control in an $N=8$ structure}
\label{sec:Wave control in an $N=8$ structure}

In Fig.~\ref{fig:figure 4}, we showed that a structure with a spatial gradient in unit-cell energy barrier heights can function as a discrete amplitude sensor, where the final configuration encodes the amplitude of the applied haversine pulse. Here, we demonstrate that the same metamaterial also has potential for wave control.

\begin{figure}
\centering
\includegraphics[width=1\linewidth]{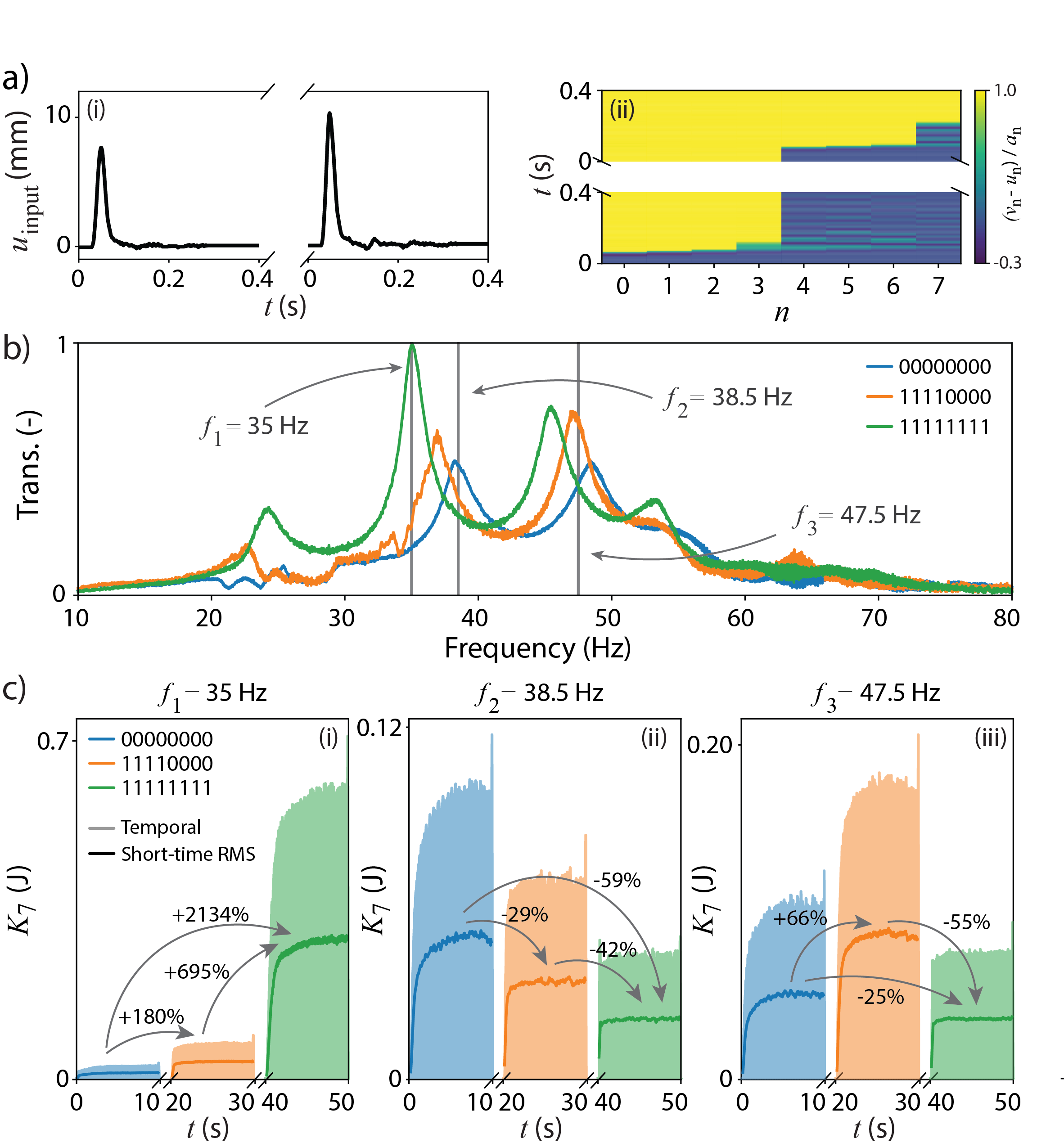}
\caption{\textbf{Wave control in an $N=8$ metamaterial.} a)(i) Experimentally recorded input signals and a)(ii) resulting spatio-temporal response of the metamaterial as it is reconfigured from state $00000000$ to state$11110000$ and then from state $11110000$ to state $11111111$. b) Configuration-dependent frequency response of $\ddot{u}_7$ normalized to the maximum of state $11111111$. c)  Configuration-dependent kinetic energy $K_7$ at three selected probe frequencies.}
\label{fig:figure 5}
\end{figure}

Guided by the results presented in Fig.~\ref{fig:figure 4}f, the metamaterial can be reconfigured from state $00000000$ to state $11110000$ by applying a single-haversine displacement pulse with frequency $\bar{f}=12$ Hz and amplitude $A=7.4$ mm, and subsequently from state $11110000$ to state $11111111$ using a second single-haversine pulse with the same frequency but an empirically determined larger amplitude of $A=10.2$ mm. Fig.~\ref{fig:figure 5}a shows the experimentally recorded input signals and the corresponding dynamic response of the sample. Discontinuities in the time axis indicate that the structure was allowed to fully relax between successive large-amplitude excitation pulses.

Once the metamaterial is resting in one of these three configurations, it is excited using a low-amplitude chirp signal with a driving frequency swept over the range $f_{\mathrm{drive}}=[8,400]$ Hz, while the acceleration of the outer mass of unit cell $n=7$, $\ddot{u}_7(t)$, is recorded. For each configuration, we compute the fast Fourier transform of $\ddot{u}_7$ to obtain frequency-dependent response amplitude of the metamaterial (Fig.~\ref{fig:figure 5}b). From these spectra, we identify three frequencies at which the dynamic response of unit cell $n=7$ depends strongly on the state of the metamaterial. Specifically, we select $f_1=35$ Hz, for which the response magnitude increases monotonically as the metamaterial transitions from state $00000000$ to $11110000$ and then to $11111111$; $f_2=38.5$ Hz, for which the response magnitude decreases monotonically across the same sequence of states; and $f_3=47.5$ Hz, for which the intermediate configuration, $11110000$, produces the largest response.

To further characterize these configuration-dependent responses, we excite the metamaterial at each of the three selected frequencies for $10$ s using low-amplitude harmonic inputs while recording $\ddot{u}_7(t)$. To isolate the response at the excitation frequency, we band-pass filter $\ddot{u}_7(t)$ using a fourth-order zero-phase Butterworth filter. Integrating the filtered acceleration signal yields the velocity, $\dot{u}_7(t)$, from which we compute the instantaneous kinetic energy,
$K_7(t)=0.5m_{\mathrm{out}}\dot{u}_7^2(t)$.
Fig.~\ref{fig:figure 5}c presents the resulting kinetic energy and its short-time root-mean-square (RMS) value as functions of time for each of the three considered configurations and excitation frequencies. At $f_1=35$ Hz, the structure acts as a reconfigurable amplifier of the transmitted signal, with $K_7$ increasing monotonically as the metamaterial is reconfigured from state $00000000$ to $11110000$ and then to $11111111$. Specifically, $K_7$ increases by $180\%$ in state $11110000$ relative to state $00000000$, followed by a further increase of $695\%$ in state $11111111$.
In contrast, at $f_2=38.5$ Hz, the structure behaves as a tunable attenuator, with $K_7$ decreasing monotonically across the same sequence of configurations. Specifically, $K_7$ decreases by $29\%$ when transitioning from state $00000000$ to $11110000$, followed by an additional decrease of $42\%$ upon transitioning to state $11111111$.
Finally, at $f_3=47.5$ Hz, the structure effectively functions as a switch, with $K_7$ first increasing by $66\%$ and subsequently decreasing by $55\%$ as the metamaterial is reconfigured from state $00000000$ to $11110000$ and then to $11111111$, respectively. While here we focus  on three representative metamaterial states, analysis of the frequency-response spectra for all $N+1$ supported states shows that, at appropriately selected excitation frequencies, the configuration-dependent dynamics can be exploited to realize $N+1$ distinct instantaneous kinetic-energy responses (Fig. S7).

\section{Conclusions}

In summary, we have demonstrated that mechanical information encoded in a mass-in-mass metamaterial can be both written and read using only pulses applied at its boundaries. Specifically, we introduced state-dependent stiffness into the bistable elements to enable non-destructive read-out through low-amplitude dynamic excitation. We further showed that this strategy extends naturally to larger arrays, enabling functional behaviors such as wave manipulation and mechanical input sensing. All together, these results demonstrate that mechanical metamaterials supporting both dynamic writing and dynamic reading can store and process information in a purely mechanical manner.

Although here we focused on one-dimensional metamaterials comprising either $N=3$ or $N=8$ unit cells, the same principles can be extended to larger arrays to increase storage capacity or to higher-dimensional architectures to introduce directional control of wave propagation. Moreover, we expect that further optimization of the unit-cell geometry could enhance both the writing and reading performance of such systems. More broadly, this work establishes a new paradigm for mechanical information processing, showing that the combination of bistability and state-dependent dynamics provides a robust foundation for programmable mechanical memory and computation.

\section{Acknowledgments}

This material is based upon research supported by the Chateaubriand Fellowship of the Office for Science $\&$ Technology of the Embassy of France in the United States. In addition, this work was supported by the Simons Collaboration on Extreme Wave Phenomena Based on Symmetries, by the National Science Foundation (NSF) under award no. 2118201 and by CNRS via IRP DynaMetaFlex.

\section{Data Availability}
The data supporting the findings of this Letter are openly available at \href{https://github.com/bertoldi-collab/dynamic_mass_in_mass_reading}{github.com/bertoldi-collab/dynamic$\_$mass$\_$in$\_$mass$\_$reading}

% \bibliography{references}

%
\end{document}